\documentclass[12pt]{article}
\usepackage{epsfig}
\newcommand{\mpl}{M_P}
\newcommand{\be}{\begin{equation}}
\newcommand{\ee}{\end{equation}}
\newcommand{\bea}{\begin{eqnarray}}
\newcommand{\eea}{\end{eqnarray}}

\title{Cosmological constraints on\\ a dark matter
-- dark energy interaction}

\author{Mark B. Hoffman \thanks{\tt mbhoffma@oddjob.uchicago.edu} 
	\\ \\
\it Enrico Fermi Institute, Department of Physics, \\
\it and Center for Cosmological Physics \\
\it University of Chicago \\
\it 5640 S.~Ellis Avenue, Chicago, IL~60637, USA }

\begin{document}
\baselineskip 20pt
\maketitle

\begin{abstract}
It is generally assumed that the two dark components of the energy
density of the universe, a smooth component called dark energy and a
fluid of nonrelativistic weakly interacting particles called dark
matter, are independent of each other and interact only through
gravity.  In this paper, we consider a class of models in which the
dark matter and dark energy interact directly.  The dark matter
particle mass is proportional to the 
value of a scalar field, and the energy density of this scalar field
comprises the dark energy.  We study the phenomenology of these models
and calculate the luminosity distance as a function of redshift and
the CMB anisotropy spectrum for several cases.  We find that the
phenomenology of these models can differ significantly from the
standard case, and current observations can already rule out the
simplest models.
\end{abstract}

\vfill\eject

\section{Introduction}

Present observations of the universe strongly suggest that roughly
ninety-six percent of the energy density of the universe is due to 
forms of matter and energy that are not described by the standard
model of particle physics.  These forms of matter and energy have
little or no direct interaction with 
ordinary matter, and hence cannot be directly observed.  The dark
sector, called such because its energy and matter do not emit light, is
typically divided into two categories, dark matter, which is
clustered, and dark energy, which is smoothly distributed and
presently causing the expansion of the universe to accelerate.  It is
also generally assumed that these two components of the dark sector
are independent and do not interact directly.  However, there are no
experiments or observations that are known to explicitly preclude such
an interaction.  The goal of this work is to conduct a preliminary
investigation into the constraints present observations place on
possible interactions in the dark sector.

Though there are presently no direct observations of the dark sector,
there are nevertheless many indirect measurements that give us
clues about its nature.  Measurements of the rotation curves of spiral
galaxies, the temperature profiles of galaxy clusters,
gravitational lensing of clusters, the large-scale motions of galaxies
between clusters, and applying the virial theorem to clusters
all require a mass for these objects much larger than that provided by
the luminous matter \cite{Persic:1995ru, Bahcall:1996mt, Primack:1997av,
Freedman:1999yu, Turner:1999kz}.  It is widely believed that the extra
mass is provided by a fluid of nonrelativistic, weakly interacting
particles known as cold dark matter (CDM).  Numerical studies of
structure formation and comparison to statistical studies of galaxies
and clusters support the CDM model on large scales \cite{Turner:1997de}. 

Observations of the anisotropy spectrum of the cosmic microwave
background radiation (CMB) \cite{Bennett:2003bz, Netterfield:2001yq,
Halverson:2001yy} and the magnitude-redshift relation for
Type Ia supernovae (SNe Ia) \cite{Riess:1998cb, Perlmutter:1998np,
Tonry:2003zg, Riess:2001gk} indicate the presence of a second
component of the dark sector, dark energy, which is smoothly
distributed and presently causing the expansion of the universe to
accelerate.  The location of the first acoustic peak of the CMB
anisotropy spectrum is predominantly governed by geometry of the
universe, and hence the total energy density of the universe.  The
measured location of the first peak implies that the universe is very
nearly flat, i.e. Euclidean.  However, the measured clustered
matter, including both ordinary matter and dark matter, only comprises
one-third of the energy density required for a flat universe implying
the existence of a smoothly distributed dark energy.  Measurements of
the magnitude-redshift relation of SNe Ia indicate that the universe
recently entered an era of accelerated expansion. The amount of dark
energy required to cause the observed acceleration also makes up for
the rest of the energy density needed to make the universe flat.

The simplest explanation for the dark energy is a cosmological
constant, $\Lambda$ (for reviews see \cite{Carroll:2000fy,
Peebles:2002gy, Padmanabhan:2002ji}).  This hypothesis fits the data
extremely well but  
faces significant theoretical questions.  There is no known mechanism
for producing a cosmological constant as small as is observed when
compared with the Planck scale ($\Lambda/\mpl^4 \sim
10^{-120}$), and this constant must be very finely
tuned to produce acceleration only very recently in the evolution of
the universe. Another possible candidate is dynamical dark energy in
which the dark energy is due to the 
potential energy of a scalar field, similar to the mechanism that is
thought to drive inflation in the early universe
\cite{Wetterich:fm,Ratra:1987rm,Frieman:1995pm,Caldwell:1997ii,
Armendariz-Picon:1999rj,Armendariz-Picon:2000dh,Armendariz-Picon:2000ah,
Mersini:2001su,Caldwell:1999ew,Carroll:2003st,Sahni:1999gb,Parker:1999td}.
A convenient
parameterization of such models is the equation of state parameter,
$w=p/\rho$, the pressure divided by the energy density of the dark
energy.  For $w\sim$ constant, the SNe Ia measurements and other
observations restrict this parameter to the range $-1.62 < w < -0.74$
\cite{Melchiorri:2002ux}, which provides a stringent constraint on
several interesting dynamical dark energy
 models.  A cosmological constant would yield $w=-1$.  If
future observations indicate $w<-1$, this would imply new physics in
the dark sector, either a rapidly varying $w$ \cite{Maor:2001ku}, a
violation of a generally accepted condition on matter called the dominant
energy condition \cite{Carroll:2003st}, or a more complicated dark
sector not yet explored. 
Moreover, though
the CDM model with a cosmological constant, known as $\Lambda$CDM,
impressively fits the current observations, new measurements in
the near future will put the $\Lambda$CDM model through strict tests.
It will be useful to have more general dark sector models around to
both act as a foil to $\Lambda$CDM and as a possible replacement
should it fail future tests.  One possibility for a more complicated
dark sector model is including an interaction between the dark matter
and the dark energy.  

Casas, Garcia-Bellido, and Quiros \cite{Casas:1992} considered models of
scalar-tensor gravity in which the scalar coupled differently to
different species of matter.  Wetterich \cite{Wetterich:bg} pointed
out that the scalar in such an interaction could constitute dark
energy.  Anderson and Carroll \cite{Anderson:1997un} studied a
cosmological model in which the dark 
matter particle mass was proportional to the expectation value of a
scalar field.  Similar models have also been studied by Bean
\cite{Bean:2001ys} and Farrar and Peebles \cite{Farrar:2003}.
Amendola introduced a model with an exponential coupling between dark
matter and a scalar field that acts as the dark energy
\cite{Amendola:1999er}.  This model has been extensively studied and
shown to be consistent with present observations
\cite{Amendola:2000uh, Amendola:2001rc, Amendola:2002bs,
Comelli:2003cv, Amendola:2003eq}.  
A particularly interesting property of this model is that it admits a
late time attractor solution with 
$\Omega_{DM}/\Omega_{DE} \sim 1/2$, the ratio observed today.  Other
possible dark sector interactions have also been studied recently 
\cite{Gasperini:2001pc, delaMacorra:2002tk, Mangano:2002gg,
Khuri:2003hf, Zimdahl:2001ar,Zimdahl:2002zb, Chimento:2003ie}.
Producing a consistent quantum theory of a dark matter -- dark energy
interaction may prove difficult \cite{Doran:2002bc}, but in principle
models could 
exist that can overcome these obstacles.  Thus, understanding the
phenomenology of such models continues to be of interest.

In this paper, we consider the interaction studied by
Anderson and Carroll \cite{Anderson:1997un} and recently by Farrar and
Peebles \cite{Farrar:2003}.  The 
model discussed here has a direct coupling between the dark matter and
dark energy.  The dark matter particle mass is proportional to the
value of a scalar field that acts as the dark energy.  One should be
concerned that by allowing the dark matter particle mass to change
with time, the dark matter energy density will be too high or too low
at earlier epochs.  This could pose significant difficulties for
structure formation and the CMB temperature anisotropy spectrum.  The
equations of motion for perturbations in these models differ, however from
those in $\Lambda$CDM and other dynamical dark energy models.  Thus,
it is conceivable that there may be new effects in this model that
compensate for the altered dark matter energy density.  Careful
calculations should be done to confirm or refute these concerns.
The
cosmological equations of motion governing this model for an arbitrary
scalar field potential are derived in Sec.~2.  In Sec.~3 we specify a
potential and explore the phenomenology of the model, which can differ
significantly from $\Lambda$CDM.  We also calculate the luminosity
distance relation for several models and compare the results to SNe Ia
data.  In Sec.~4 we consider
perturbations in this model and calculate the CMB anisotropy spectrum
for several cases.  We conclude in Sec.~5 with a brief discussion of
the results and future prospects for interacting dark sector models.

\section{Dark Sector Equations of Motion}

We begin this section with an intuitive motivation for the equations
of motion governing a fluid of dark matter particles with a rest mass
that is proportional to the value of a scalar field.  A more detailed
derivation from an action for two scalar fields follows.  We consider
here a model that is a flat, 3+1 dimensional
Friedmann-Robertson-Walker (FRW) universe with metric 
\be
\label{metric}
ds^2 = -dt^2 + a^2\delta_{ij}dx^idx^j.
\ee

The dark matter particles in this model will be similar to the
particles of the CDM model: they will be collisionless and
nonrelativisic. Hence, the pressure of this fluid vanishes and the
energy density is given by the rest mass, $m$, multiplied by the
number density, $n$, of the dark matter particles.  We define 
\[
m = \lambda\phi,
\]
where $\lambda$ is a dimensionless constant and $\phi$ is a scalar
field.  The energy density and pressure associated with this fluid are
thus
\bea
\rho_{DM} &=& mn = \lambda\phi n \nonumber \\
p_{DM} &=& 0. \nonumber
\eea
We will
assume that this species of particle froze out in the early universe
so that the comoving number density of dark matter particles is
constant during the epochs of interest, i.e the particles are neither
created nor destroyed.  Thus, the number density is only a function of
physical volume and $n=n_0 a^{-3}$ where $n_0$ is the present number
density of dark matter particles.  

One may be surprised that the
dark matter is described by a pressureless fluid for which $\rho \neq
\rho_0(a/a_0)^{-3}$.  The result $\rho = \rho_0(a/a_0)^{-3}$ for a
pressureless fluid is due to the vanishing of the divergence of the
stress-energy of a {\em non-interacting} pressureless fluid.  The dark
matter studied in this work interacts directly with a scalar field,
and thus its stress-energy does not have vanishing divergence.  We
derive the evolution equation for the energy density of the
interacting dark matter below.

The scalar field that defines the mass of the dark matter particles
has a standard kinetic term and a potential
$V(\phi)$.  The energy density and pressure associated with the scalar
field are
\bea
\rho_\phi &=& \frac{1}{2}\dot{\phi}^2 + V(\phi) \nonumber \\
p_\phi &=& \frac{1}{2}\dot{\phi}^2 - V(\phi), \nonumber
\eea
where an overdot represents a derivative with respect to time, $d/dt$.
Since the energy density of the dark matter particles depends on
$\phi$, the scalar field feels an additional effective potential when
it is in a bath of dark matter particles.  Taking this effect
into account, the equation of motion is
\be
\label{eq:EoM_phi_gen}
\ddot{\phi}+3H\dot{\phi}+\frac{dV_{\rm eff}}{d\phi} = 0,
\ee
where $H = \dot{a}/a$ and 
\[
V_{\rm eff}(\phi) = V(\phi) + \lambda n \phi.
\]
Substituting this back into (\ref{eq:EoM_phi_gen}) yields
\be
\label{eq:EoM_phi}
\ddot{\phi}+3H\dot{\phi}+\frac{dV}{d\phi} = -\lambda n_0a^{-3}.
\ee
The only difference between this equation and that for a
noninteracting dynamical dark energy model is the term on the right hand side,
which accounts for the interaction.

In order to derive an evolution equation for the dark matter energy
density, we first consider the divergence of the stress-energy tensor
for each dark component.  Since neither dark component interacts
directly with any other species, the divergence of the sum of their
stress-energy tensors must vanish.  However, due to the interaction,
the divergence of each stress-energy tensor is not necessarily zero.
The derivative operator is linear, so
\[
\nabla_\mu\left(T_{(DM)}{}^\mu{}_\nu+T_{(\phi)}{}^\mu{}_\nu\right)
	=\nabla_\mu T_{(DM)}{}^\mu{}_\nu+\nabla_\mu T_{(\phi)}{}^\mu{}_\nu,
	= 0,
\]
which implies
\be
\label{eq:div_T}
\nabla_\mu T_{(DM)}{}^\mu{}_\nu = -\nabla_\mu T_{(\phi)}{}^\mu{}_\nu.
\ee
The stress-energy tensor for the dark matter, $T_{(DM)}{}^\mu{}_\nu$, is
fairly simple.  The only non-vanishing component is $T_{(DM)}{}^0{}_0 =
-\rho_{DM}$.  For the scalar field, the stress-energy tensor is
\[
T_{(\phi)}{}^\mu{}_\nu = \partial^\mu\phi\partial_\nu\phi
	- \delta^\mu{}_\nu\left[
	\frac{1}{2}\partial^\alpha\phi\partial_\alpha\phi +
	V(\phi)\right],
\]
and its divergence is
\bea
\nabla_\mu T_{(\phi)}{}^\mu{}_\nu &=& \partial_\mu
	T_{(\phi)}{}^\mu{}_\nu 
	+ \Gamma^\mu_{\mu\beta}T_{(\phi)}{}^\beta{}_\nu
	- \Gamma^\beta_{\mu\nu}T_{(\phi)}{}^\mu{}_\beta \nonumber \\
 &=& -\left(\ddot{\phi}+3H\dot{\phi}+\frac{dV}{d\phi}\right)
	\partial_\nu\phi .
\eea
Using the equation of motion for the scalar field (\ref{eq:EoM_phi}), this
expression simplifies to
\be
\label{eq:div_phi}
\nabla_\mu T_{(\phi)}{}^\mu{}_\nu = \lambda n\partial_\nu\phi.
\ee
The evolution equation for the dark matter energy density is then
calculated by combining (\ref{eq:div_T}) and (\ref{eq:div_phi}):
\be
\label{eq:div_DM}
-\nabla_\mu T_{(DM)}{}^\mu{}_0 = \dot{\rho}_{DM}+3\frac{\dot{a}}{a}\rho
	= \lambda n \dot{\phi}.
\ee
As a consistency check, we can use this equation and the definition of
$\rho_{DM}$ to determine the evolution of the number density as
a function of the scale factor:
\[
\dot{\rho}_{DM} = \lambda\dot{n}\phi + \lambda n\dot{\phi}
	+3\frac{\dot{a}}{a}\lambda n\phi = \lambda n\dot{\phi},
\]
which trivially reduces to
\[
\dot{n} = -3\frac{\dot{a}}{a}n
\]
and is solved by $n = n_0 (a/a_0)^{-3}$, as we had earlier assumed.

The scale factor of the universe evolves according to the Friedmann
equation 
\be
\label{eq:FRW}
\left(\frac{\dot{a}}{a}\right)^2 = \frac{8\pi G}{3}\rho,
\ee
where $\rho$ is the total energy density of the universe and $G$ is
Newton's constant.  In the
models considered in this paper, the universe contains the dark matter
and  dark energy discussed so far along with baryons, whose energy
density scales as $(a/a_0)^{-3}$, and radiation, whose energy density
scales as $(a/a_0)^{-4}$.  Thus, the energy density as a function of
the scale factor is:
\[
\rho = \rho_{DM} + \rho_\phi + \rho_{B,0}(a/a_0)^{-3}
	+ \rho_{R,0}(a/a_0)^{-4}.
\]

We now continue with a more explicit derivation of these equations
from an action for two interacting scalar fields, one of which will
play the role of dark energy and the other dark matter. 
Consider a scalar dark matter particle, $\psi$, and a homogeneous
scalar field, $\phi$, which serves as the source of the 
dark energy.  The action for this model is
\be
\label{action}
S = \int d^4x \,\sqrt{-g}\left[\frac{1}{2}\mpl^2R
	-\frac{1}{2}\partial^\mu\phi\partial_\mu\phi - V(\phi)
	-\frac{1}{2}\partial^\mu\psi\partial_\mu\psi
	-\frac{1}{2}\lambda^2\phi^2\psi^2\right],
\ee
where $R$ is the Ricci scalar associated with metric (\ref{metric}),
$g$ is the determinant of the metric, and $\mpl = 1/\sqrt{8\pi G}$
is the reduced Planck mass.  Varying this action with respect to the
scalar fields yields 
\bea
\label{phi_EOM}
\ddot{\phi}+3H\dot{\phi}+\frac{dV}{d\phi}+\lambda^2\phi\psi^2 = 0\\
\label{psi_EOM}
\ddot{\psi}-\nabla^2\psi+3H\dot{\psi}+\lambda^2\phi^2\psi = 0.
\eea
The spatial gradient term vanishes in
the equation for $\phi$ because this field, which will make up the
dark energy, is assumed to be spatially homogeneous.

We now consider the equation of motion for the $\psi$ particle.  The
time scale associated with the dynamics of a particle of mass $m$ is
$\tau\sim m^{-1}$.  To simplify the calculation, we will 
assume that $m_\psi \gg H$, which is not a restrictive constraint
during the epochs we will be considering.  The mass of the $\psi$
field, $m_\psi\equiv\lambda\phi$, evolves over short times as
\bea
m_\psi &\approx& m_{\psi,*} + \frac{dm_\psi}{dt}dt \nonumber \\
 &\approx& m_{\psi,*}\left(1+\frac{\dot{m_\psi}}{m_{\psi,*}}dt\right)
	\nonumber \\
 &\approx& m_{\psi,*}\left(1+\frac{\dot{\phi}}{\phi_*}dt\right), \nonumber
\eea
where $m_{\psi,*}$ is the $\psi$ mass at some fixed time, $\phi_*$ is
the value of the $\phi$ field at the same fixed time, and $dt$ is a
time interval of order the time scale of the evolution of the $\psi$
particle. We expect that the $\phi$ field, which is the dark energy in
this model, evolves on a time scale that is comparable to the time scale
associated with the evolution of the universe, i.e. $\dot{\phi}/\phi
\sim H$.  Using this assumption and setting $dt \sim \tau \sim
m_\psi^{-1}$ yields
\[
m_\psi \approx  m_{\psi,*}\left(1+\frac{H}{m_\psi}\right).
\]
Since we have assumed that $H/m_\psi \ll 1$, the mass of the $\psi$
particle is a constant to a very good approximation.  This reasoning
also implies that the Hubble term in Eq. (\ref{psi_EOM}) is
negligible.  This is reasonable because on time scales much shorter
than that associated with the expansion of the universe, the expansion
should be negligible.  

The equation of motion for the $\psi$
particle becomes after these approximations
\[
\ddot{\psi}-\nabla^2\psi+m_\psi^2\psi = 0,
\]
which not surprisingly is just the Minkowski space Klein-Gordon
equation.  The only difference here is that over long times, i.e. on
cosmological scales, $m_\psi$ changes.  

On timescales much smaller than the Hubble time, we can ignore the
expansion of the universe and use the standard results from quantum
field theory in Minkowski spacetime to study the properties of the
particles associated with the $\psi$ field.  We will work in the
Schr\"{o}dinger picture in which the field operators are time
independent.  We begin by decomposing the $\psi$ field into its
Fourier modes: 
\[
\psi({\bf x}) = \int\frac{d^3p}{(2\pi)^3}\frac{1}{2\omega_{\bf p}}
	\left(a_{\bf p}e^{i{\bf p}\cdot{\bf x}}
		+a_{\bf p}^\dagger e^{-i{\bf p}\cdot{\bf x}}\right),
\]
where boldface indicates a 3-vector.
The energy associated with this field is given by the associated
(renormalized) Hamiltonian 
\be
\label{hamiltonian}
{\mathcal H} = \int \frac{d^3p}{(2\pi)^3}\omega_{\bf p}
	\left(a_{\bf p}^\dagger a_{\bf p}\right),
\ee
where $\omega_{\bf p} = \sqrt{|{\bf p}|^2 + m_\psi^2}$ and $a_{\bf
p}^\dagger$ and $a_{\bf p}$ are the usual creation and annihilation
operators.  Since $a_{\bf p}^\dagger a_{\bf p}$ acting on a state
yields the number of particles with momentum ${\bf p}$, we can interpret
(\ref{hamiltonian}) as
\[
E = \int d^3p\, N_{\bf p}\omega_{\bf p}.
\]
Recalling that the dark matter particles are assumed to be
nonrelativistic, i.e. the distribution of $\psi$ particles is
dominated by particles with $|{\bf p}| \ll m_\psi$, then it follows 
that the energy density in dark matter particles is given by
\[
\rho_\psi = m_\psi n_\psi = \lambda\phi n_\psi.
\]

We now turn our attention to (\ref{phi_EOM}), the equation of motion
for the $\phi$ field, which we will treat classically.  To do this, we
need to interpret the $\lambda^2\phi\psi^2$ term.  Since we are
treating the $\phi$ field classically, we will consider the term:
$\lambda^2\phi\left<n|\psi^2|n\right>$ where
$\left<n|\psi^2|n\right>$ is the expectation value of $\psi$ in
a state with $n$ particles per unit volume.  To evaluate the $\psi$
term, we begin by considering the $\psi^2$ operator:
\bea
\psi^2({\bf x}) &=& \int\frac{d^3p}{(2\pi)^3}\frac{d^3p^\prime}{(2\pi)^3}
	e^{i({\bf p+p^\prime})\cdot{\bf x}}
	\frac{1}{2\sqrt{\omega_{\bf p}\omega_{\bf p^\prime}}}
	\left(a_{\bf p}+a_{-\bf p}^\dagger\right)
	\left(a_{\bf p^\prime}+a_{-\bf p^\prime}^\dagger\right) \nonumber \\
 &=& \int\frac{d^3p}{(2\pi)^3} \frac{1}{2\omega_{\bf p}}
	\left(a_{\bf p}a_{\bf -p}
		+ a_{\bf p}^\dagger a_{\bf -p}^\dagger
		+ a_{\bf -p}^\dagger a_{\bf -p} 
		+ a_{\bf p}^\dagger a_{\bf p}
		+ \left[a_{\bf p},a_{\bf p}^\dagger\right]\right) \nonumber.
\eea
Since we are assuming that the $\psi$ particles are non-relativistic
and it will greatly simplify the calculation, let us assume that all
of the particles are in the ${\bf p} = 0$ state.  Then we have the
(renormalized) result:
\[
\left<n|\psi^2|n\right> = \frac{n}{\omega_{{\bf p}=0}}
	= \frac{n}{m_\psi}
	= \frac{n}{\lambda\phi}
\]
Substituting this back into (\ref{phi_EOM}) yields
\be
\label{EOM}
\ddot{\phi}+3H\dot{\phi}+\frac{dV}{d\phi}+\lambda n = 0,
\ee
which is precisely the equation of motion motivated earlier.  The rest
of the classical discussion proceeds as already discussed in this section.
Though we derived this equation for a scalar particle, it holds
equally well for nonrelativistic fermions as shown in \cite{Farrar:2003}.

\section{Cosmology with an inverse power law potential}

Exploring the phenomenology of the model
described in the previous section requires a form for the potential of
the scalar field, $V(\phi)$.  Since 
the mass of the dark matter particle depends linearly on $\phi$, we
choose a potential that blows up as $\phi$ approaches zero and thus
prevents $\phi$ from becoming negative:
\be
\label{eq:Vphi}
V(\phi) = K\phi^{-\alpha}.
\ee
This form for the potential is also used in \cite{Anderson:1997un} and
\cite{Farrar:2003}.  While this form of the potential is not necessary, 
the features of the cosmology resulting from this choice are
significant, making it a good illustrative example.
Substituting the potential (\ref{eq:Vphi}) into the field equation
(\ref{eq:EoM_phi}) yields 
\be
\label{eq:EoM_V}
\ddot{\phi}+3H\dot{\phi}-\alpha K\phi^{-\alpha-1}+\lambda n_0a^{-3}=0.
\ee

\begin{figure}
\begin{center}
\epsfig{file=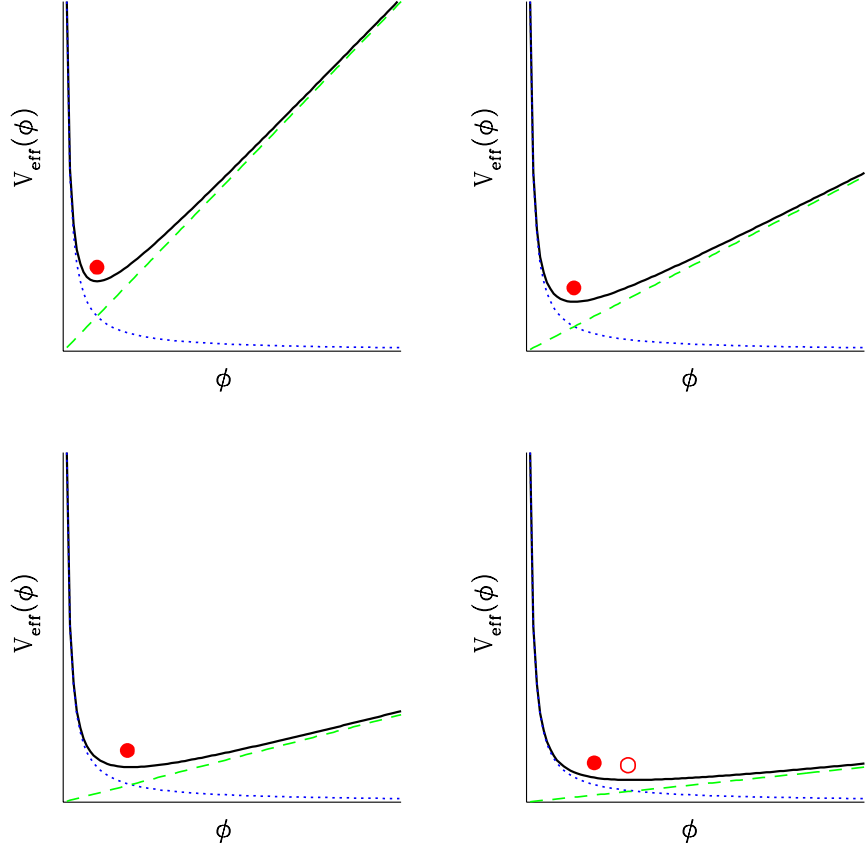, width=5in}
\end{center}
\caption{\label{Veff}
The effective potential, $V_{\rm eff}(\phi)$ changes as the
universe expands due to the term that depends on the number density of
dark matter particles.  In all four pictures, the dotted line is
the scalar field potential, the dashed line is the rest energy
of the dark matter particles as a function of $\phi$, and the 
solid line is the effective potential.  In the early universe (top
left), the number density of the dark matter particles is large
causing the line associated with it to be steep.  The field rapidly
settles to the minimum of the effective potential.  As the universe
evolves (top right), this line becomes less steep, and
the minimum moves to larger $\phi$.  The field (solid circle)
follows the minimum until (lower left), the $3H\dot{\phi}$ term
in the field equation (\ref{eq:EoM_phi})
becomes comparable to the first derivative of the potential.  After
this time, the $3H\dot{\phi}$ term acts as an effective friction and slows
the field.  Eventually the field is no longer in the minimum of the
potential and is moving very slowly (lower right).  The open circle in
the lower right plot shows the minimum of the potential, and the solid
circle shows the value of the field.  At this point, the field is
close to constant, and the interacting dark matter starts to behave as
standard cold dark matter with a fixed mass.}
\end{figure}

In the early universe, the effective potential looks like the upper
left picture in Fig.~\ref{Veff}.  The effective potential is
very steep on both sides of the minimum, and the minimum is at a small
value of $\phi$.  Due to the steepness of the effective
potential and the effective friction due to the
$3H\dot{\phi}$ term in (\ref{eq:EoM_V}), the field
rapidly settles into the minimum of the effective potential for a wide
range of initial conditions.  As the universe expands, the number
density of dark matter particles decreases, and the minimum of the
effective potential moves to larger values of $\phi$ as seen in
the upper right picture in Fig.~\ref{Veff}.  At the same time,
the effective potential becomes shallower, and eventually the
effective friction becomes important in the evolution of the scalar
field.  This slows the field so that it cannot keep up with the
shifting minimum of the effective potential.  Eventually, the field is
far from the minimum and effectively constant.  This is shown in the
lower right picture of Fig.~\ref{Veff}.  The solid circle
shows the actual field value, and the outlined circle shows the
location of the minimum of the effective potential.  When the field
slows down sufficiently, the interacting dark matter particles begin
to act just like ordinary dark matter particles with fixed mass and
the field becomes like an ordinary non-interacting dynamical dark
energy field. 

\begin{figure}
\begin{center}
\epsfig{file=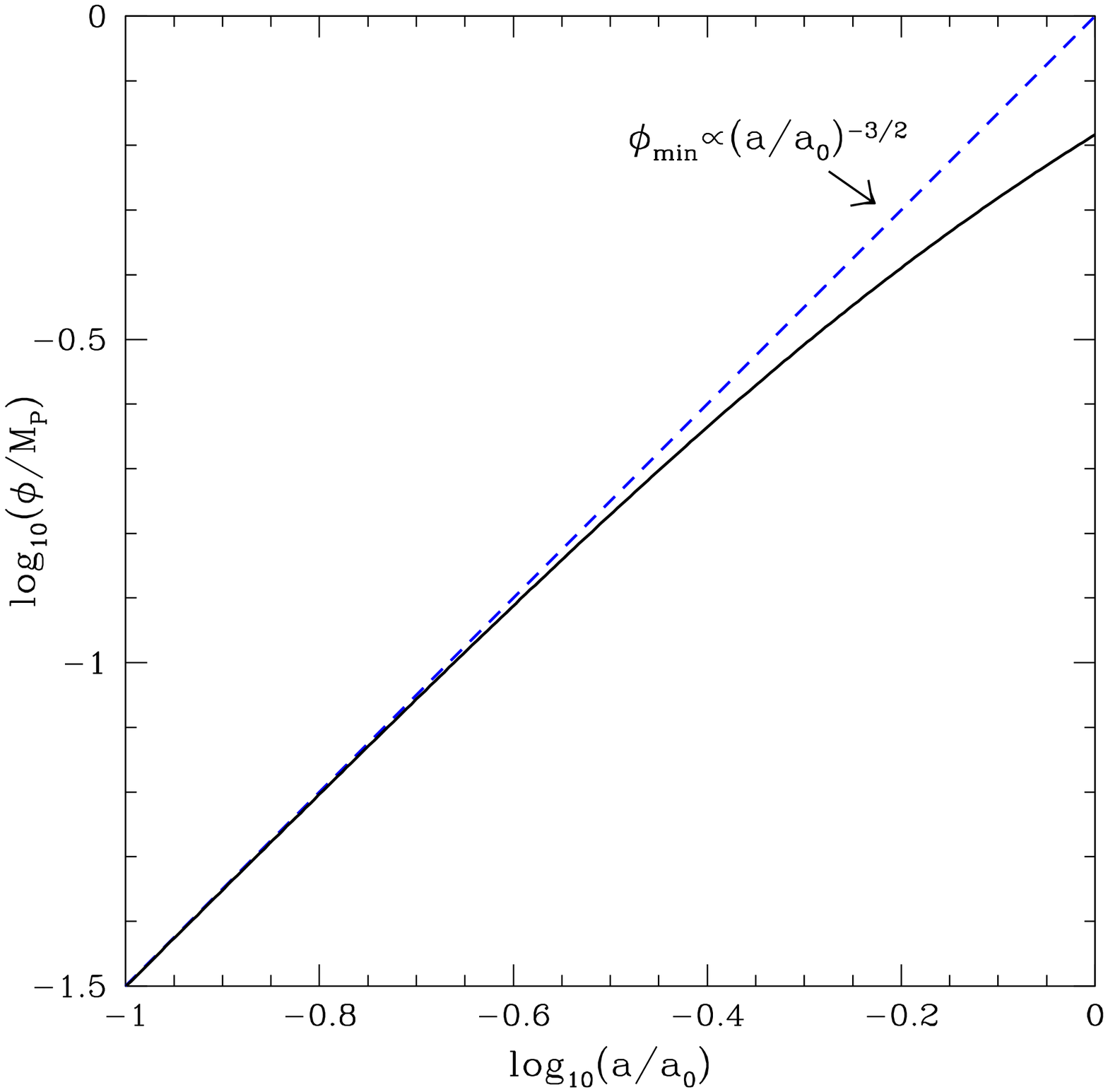, width=5in}
\end{center}
\caption{\label{phi}
Plot of $\log_{10}(\phi/\mpl)$ versus $\log_{10}(a/a_0)$
for $\alpha = 1$.  The dashed line shows the
value of the field at the minimum of the effective potential,
$\phi_{\rm min}/\mpl$ as a
function of the scale factor.  The solid line represents the actual
evolution of the field.  In the early universe, the field sits at the
minimum of the effective potential, but eventually the effective
friction due to the expansion of the universe causes the field to slow
down and fall behind the shifting minimum of the effective potential.}
\end{figure}

The evolution of the field when the field is in the minimum of the
effective potential was first calculated in \cite{Anderson:1997un}.  We
reproduce and expand upon those results here.  The value for the field
at the minimum of the effective potential is obtained by solving for
the field value that causes the first derivative of the effective
potential to vanish, and the result is
\be
\label{eq:phimin}
\phi_{\rm min} = \left(\frac{\alpha K}{\lambda n_0}\right)
	^{(1/1+\alpha)}\left(\frac{a}{a_0}\right)^{3/(1+\alpha)} 
	= \phi_0\left(\frac{a}{a_0}\right)^{3/(1+\alpha)} .
\ee
In Fig.~\ref{phi} we plot the value of the scalar field as a
function of $a/a_0$ for the case $\alpha = 1$.  In the early
universe, the field takes the value
$\phi_{\rm min}$.  At $\log_{10}(a/a_0) \sim -0.6$, the
effective friction term begins to become important and slows
the field down.  This is seen in the plot as the actual value
of the field falls below $\phi_{\rm min}$ for
$\log_{10}(a/a_0) > -0.6$.

While the field remains in the minimum of the effective potential, the
mass of the dark matter particles is  
\[
m_{DM} = \lambda\left(\frac{\alpha K}
	{\lambda\bar{n}_0}\right)^{1/(1+\alpha)}
	\left(\frac{a}{a_0}\right)^{3/(1+\alpha)},
\]
and the energy density of the dark matter is
\be
\label{eq:rhodm}
\rho_{DM} = \lambda\bar{n}_0\left(\frac{\alpha K}
	{\lambda n_0}\right)^{1/(1+\alpha)}
	\left(\frac{a}{a_0}\right)^{-3\alpha/(1+\alpha)}.
\ee
We next consider the energy density associated with the scalar field.
When the field is at the minimum of its effective potential, and hence
following (\ref{eq:phimin}), its time derivative is
\[
\dot{\phi} = \frac{3}{1+\alpha}H\phi.
\]
The magnitude of the ratio of the kinetic to potential energy of the
field is then
\be
\label{eq:NRG_ratio}
\frac{\dot{\phi}^2}{V} \sim
	\frac{H^2\phi^2}{\phi^{-\alpha}}.
\ee
Recall, however, that in order for the field to stay in the minimum of
the effective potential, the first derivative of the effective
potential must be much larger than the effective friction term in the
equation of motion (\ref{eq:EoM_phi}) implying
\be
\label{eq:eq_cond}
\frac{3H\dot{\phi}}{dV/d\phi} \sim
	\frac{H^2\phi}{\phi^{-\alpha-1}}
	\ll 1.
\ee
Comparing this to (\ref{eq:NRG_ratio}), we see that when the field is in
the minimum of the effective potential, the kinetic energy is much
smaller than the potential energy of the scalar field.  Thus to a
good approximation, the energy density in the scalar field is
\be 
\label{eq:rhophi}
\rho_{\phi} = V(\phi) = 
	K\left(\frac{\alpha K}{\lambda n_0}\right)
	^{(-\alpha/1+\alpha)}
	\left(\frac{a}{a_0}\right)^{-3\alpha/(1+\alpha)}.
\ee
We see from (\ref{eq:rhodm}) and (\ref{eq:rhophi}) that the dark
matter energy density and the energy density of the scalar field have
a constant ratio,
\[
\frac{\rho_{DM}}{\rho_{\phi}} = \alpha,
\]
while the field is in the minimum of the effective potential.
Substituting (\ref{eq:phimin}) into (\ref{eq:eq_cond}), we find that
the approximation that the field is in the minimum of the effective
potential breaks down when
\[
\frac{a_*}{a_o} 
	\sim H^{-\frac{2}{3}\left(\frac{\alpha+1}{\alpha+2}\right)}
		\left(\frac{\lambda n_0}{\alpha K}\right).
\]

\begin{figure}
\begin{center}
\epsfig{file=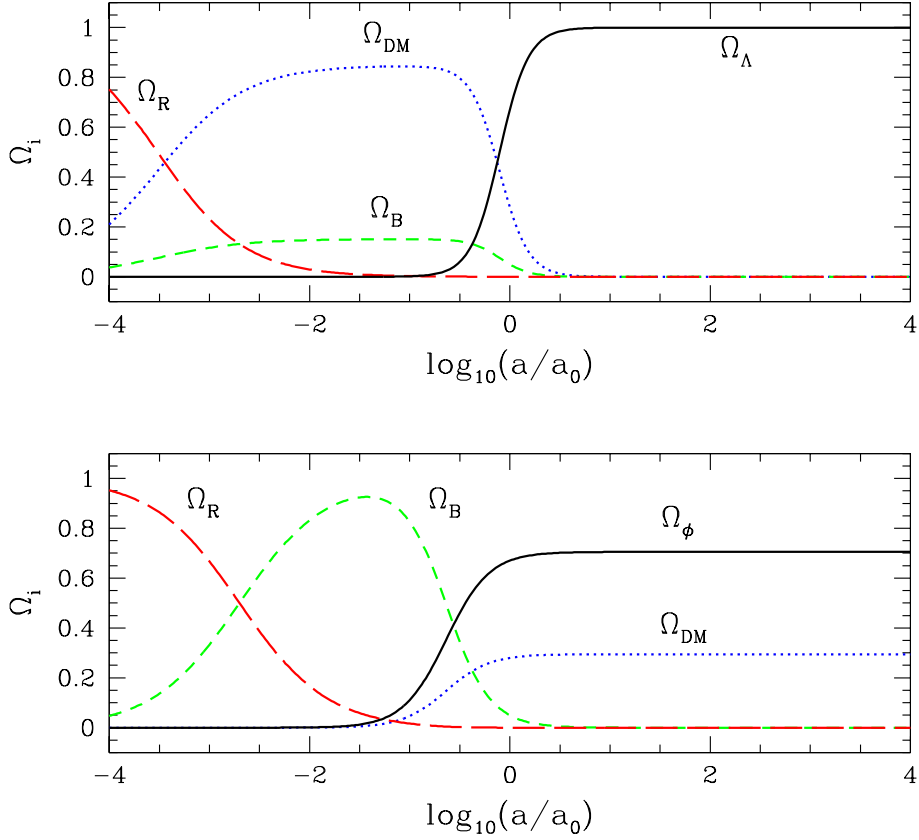, width=5in}
\end{center}
\caption{\label{omega_a.5}
The bottom panel is a plot of the relative energy density in
radiation ($\Omega_R$, long-dashed line), baryons ($\Omega_B$,
short-dashed line), dark matter ($\Omega_{DM}$,dotted
line), and the scalar field acting as the dark energy ($\Omega_\phi$,
solid line) for $\alpha = 1/2$ and $(\lambda n_0/\alpha K) \gg 1$.
The top 
panel is the same plot for $\Lambda$CDM with $\Omega_\Lambda$, the
relative energy density in the cosmological constant, replacing
$\Omega_\phi$ and is included for comparison. Eventually, the dark
energy will dominate in the interacting model, but this can be put off
as long as desired by increasing $(\lambda n_0/\alpha K)$.  Note that in
this case, $\Omega_{DM}/\Omega_\phi = 1/2$ is sustained for many
e-foldings of the scale factor.  Another feature of this model is that
the onset of dark energy density takes place over a much longer time
than in the $\Lambda$CDM scenario.  Finally, the universe is never
dominated by dark matter, but instead goes through an epoch of baryon
domination.}
\end{figure}

An interesting parameter choice for this model is $\alpha = 1/2$ and
$(\lambda n_0/\alpha K) \gg 1$.  In this case, there is an extended period
of time during which $\rho_{DM}/\rho_{\phi} = 1/2$ as is observed.  We
plot the relative energy densities of radiation ($\Omega_R$), baryons
($\Omega_B$), dark matter ($\Omega_{DM}$), and the scalar field
($\Omega_\phi$) as a function of $a/a_0$ in the bottom panel of
Fig.~\ref{omega_a.5}.  In the top panel, we plot the relative
energy densities for a $\Lambda$CDM model for comparison.

\begin{figure}
\begin{center}
\epsfig{file=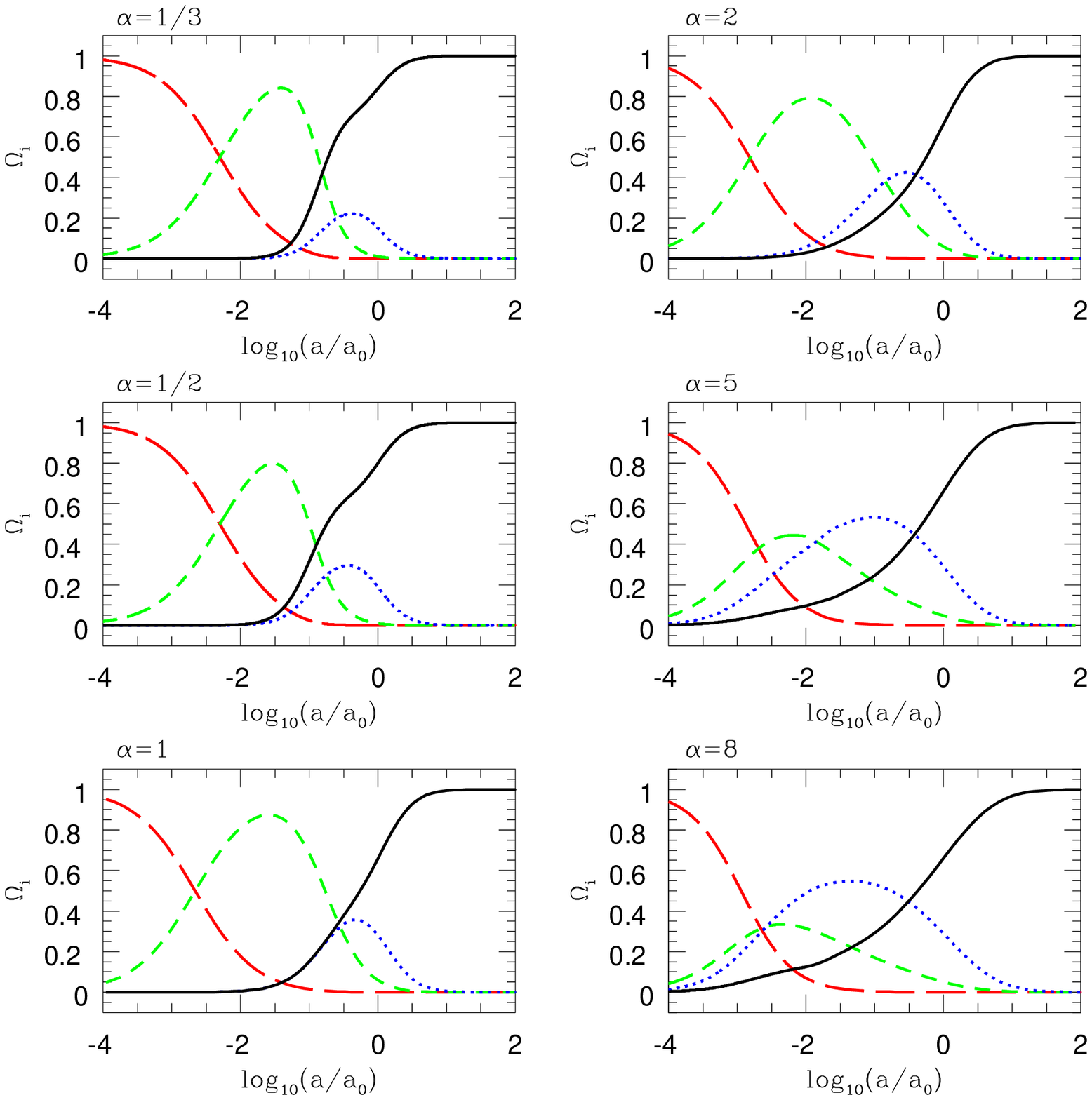, width=5in}
\end{center}
\caption{\label{omega}
Plots of the relative energy density in radiation
($\Omega_R$, long-dashed line), baryons ($\Omega_B$,
short-dashed line), dark matter ($\Omega_{DM}$, dotted line), and
the scalar field acting as the dark energy ($\Omega_\phi$, solid
line) vs. $log_{10}(a/a_0)$ for various values of $\alpha$.} 
\end{figure}

For any other value of $\alpha$, we require $a_*/a_0 < 1$ in order to
have $\rho_{DM}/\rho_{\phi} \sim 2$ today.  In Fig.~\ref{omega} we plot
the relative energy densities in a number of models with different
values of $\alpha$ and fixed $\lambda n_0$.  Changing
$\lambda n_0$ does not significantly alter these plots, since we
must adjust the value of $K$ in order to get the right energy density in
baryons and radiation today.

It is typical in models with small values of $\alpha$ that the
universe is baryon dominated during structure formation, which would
likely produce structure that is significantly different from what we
observe.  Calculating the formation of structure in these interacting
dark sector models is beyond the scope of this paper.  Instead, we
will compare these models to observations in two other areas, the
magnitude-redshift relation of SNe Ia in this section and the CMB
anisotropy spectrum in the next.

Often in comparing dynamical dark energy models to SNe Ia data, one
calculates the dark energy equation of state parameter, $w$, and
compares this theoretical value to the values for $w$ = constant dynamical
dark energy models allowed by the data.  Such a comparison will not
work in this case, however, because the $w$ = constant dynamical dark
energy models assume that the dark matter energy density  scales as
$(a/a_0)^{-3}$, 
which is not the case in interacting dark sector models.  Thus, we
must calculate the predicted magnitude-redshift relation for SNe Ia
for each model and compare the predicted relation to the measured one.

The magnitude-redshift relation of SNe Ia measures the luminosity
distance, $d_L$, as a function of redshift, $z$, via the relation
\[
m-M = \log_{10}d_L(z),
\]
where $m$ is the relative magnitude of the supernova and $M$ is
its absolute magnitude.  Typically, the results
of these measurements are plotted as the difference between the
observed $(m-M)$ and that expected from a standard model called the
``empty'' universe:
\[
\Delta(m-M) = \log_{10}d_L(z) - \log_{10}d_L^{\, \rm empty}(z).
\]
The luminosity distance is computed from the evolution of the scale
factor through the relation
\[
d_L(z) = (1+z)\int_0^z\frac{dz^\prime}{H(z^\prime)},
\]
and the result for the ``empty'' universe is 
\[
d_L^{\, \rm empty}(z) = \frac{z(1+z/2)}{H_0}.
\]

\begin{figure}
\begin{center}
\epsfig{file=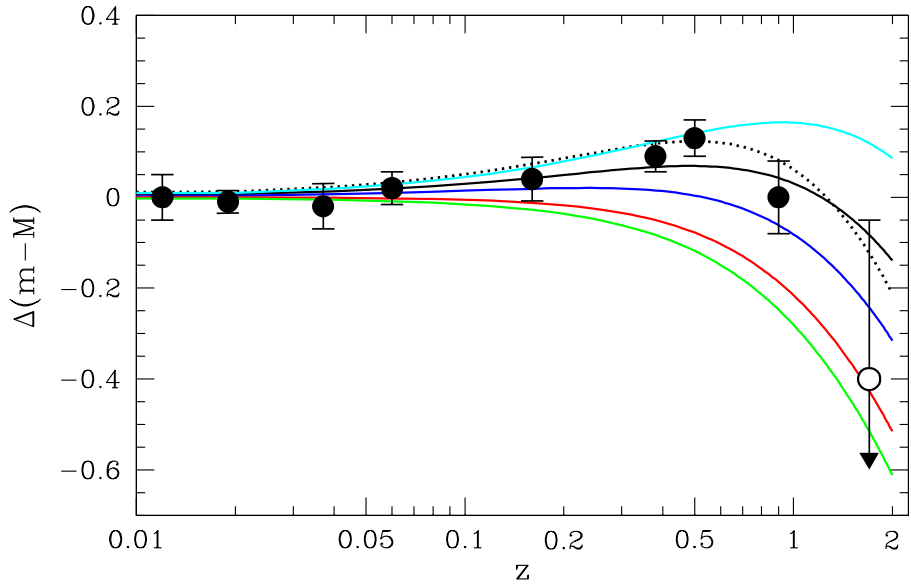, width=5in}
\end{center}
\caption{\label{dlz}
Comparison of several models with SNe Ia data.  The solid lines
lines represent models with (from top to bottom) $\alpha =
1/2,1,2,5,8$.  The theoretical curve for $\Lambda$CDM is shown as a
dotted line for reference.  The solid circles are binned SNe Ia data
found in \cite{Tonry:2003zg}, and the point at $z=1.7$ is the upper limit
for SN1997ff from \cite{Riess:2001gk}.  The models
with $\alpha > 2$ are disfavored by moderate redshift
observations.}
\end{figure}

We have calculated the theoretical $\Delta(m-M)$ for a number of
models, and the results are plotted in Fig.~\ref{dlz} along with
binned data.  For comparison, we have also plotted the $\Lambda$CDM
result. Note that large values of $\alpha$ do not fit the data well
because they do not have enough acceleration in the expansion of the
universe to match the data.  Small values of $\alpha$ may also be
problematic because acceleration begins too early to match the result
of SN1997ff at $z\sim 1.7$.  However, further high-redshift data will
be required to strictly rule out such models.  Similar results have
also been obtained for the models considered by Amendola and his group
\cite{Amendola:2002kd}. 

\section{Perturbations and the CMB anisotropy spectrum}

In this section we derive the equations of motion for perturbations of
the scalar field and the dark matter particles, which will be used in
calculating the CMB anisotropy spectrum.  We will work in conformal
time ($ad\tau = dt$) and in the synchronous gauge where the metric is 
\[
ds^2 = a^2(\tau)\left[-d\tau^2 
	+ \left(\delta_{ij}+h_{ij}\right)dx^idx^j\right].
\]
The scalar field and the dark matter energy density can be written as
the sum of a background, average value and a perturbation that is
assumed to be small when compared with the background value:
\[
\phi = \bar{\phi} + \delta\phi
\]
\[
\rho_{DM} = \bar{\rho}_{DM} + \delta\rho_{DM}.
\]
The dark matter perturbation has two components, the perturbation in
the number density of the dark matter particles, $\delta n$, and the
perturbation in the scalar field, $\delta\phi$:
\[
\delta\rho_{DM} = \lambda\bar{\phi}\delta n
	+ \lambda \bar{n}\delta\phi,
\]
where $\bar{n}$ is the average number density of dark matter particles.
It will be useful to also define the relative perturbation in the
number density of the dark matter particles:
\[
\delta_n = \frac{\delta n}{\bar{n}}.
\]

The equations of motion for the background are (\ref{eq:EoM_phi}),
(\ref{eq:div_DM}), and (\ref{eq:FRW}), which we reproduce here in terms
of the conformal time, $\tau$:
\be
\label{eq:EoM_cphi}
\bar{\phi}^{\prime\prime}+2\frac{a^\prime}{a}\bar{\phi}^\prime
	+a^2\frac{dV}{d\bar{\phi}} = -\lambda n_0/a
\ee
\be
\label{eq:EoM_cDM}
\bar{\rho}^\prime_{DM}+3\frac{a^\prime}{a}\bar{\rho}_{DM}
	= \lambda \bar{n} \bar{\phi}^\prime
\ee
\be
\label{eq:cFRW}
\left(\frac{a^\prime}{a}\right)^2 = \frac{\rho a^2}{3\mpl^2},
\ee
where a prime indicates a derivative with respect to the conformal
time, $d/d\tau$. The total scalar field satisfies the equation
\[
\nabla^\mu\nabla_\mu\phi + \frac{dV}{d\phi} 
	+ \lambda n = 0.
\]
A Fourier mode of the scalar field perturbation
\[
\delta\phi_k(\tau) = \frac{1}{\sqrt{2\pi}}
	\int \delta\phi(\tau,{\bf x})e^{-i{\bf k}\cdot{\bf x}}d^3x
\]
satisfies the equation of motion
\be
\label{eq:EoM_dphi}
\delta\phi^{\prime\prime}_k + 2\frac{a^\prime}{a}\delta\phi^\prime_k
	+ \left(k^2 + a^2\frac{d^2V}{d\phi^2}\right)\delta\phi_k
	+ \lambda \bar{n} a^2 \delta_n 
	= -\frac{1}{2}h^\prime\bar{\phi}^\prime,
\ee
where we have used the background equation of motion
(\ref{eq:EoM_cphi}) and $h$ is the trace of the metric perturbation
$h_{ij}$.  From here on we will suppress the subscript $k$ on the
perturbed variables.

In order to get the equation of motion for the dark matter
perturbation, we will need to consider the divergence of the perturbed
stress-energy tensor.  Let us begin by considering the perturbed
stress-energy tensor for the scalar field,
\[
\delta T_{(\phi)}{}^{\mu}{}_\nu 
	= \left(\partial^\mu\delta\phi\right)
		\left(\partial_\nu\bar{\phi}\right) 
	+ \left(\partial^\mu\bar{\phi}\right)
		\left(\partial_\nu\delta\phi\right)
	- \delta^\mu{}_\nu\left(
	  \frac{1}{2}\left(\partial_\alpha\delta\phi\right)
		\left(\partial^\alpha\bar{\phi}\right)
	  + \frac{1}{2}\left(\partial_\alpha\bar{\phi}\right)
		\left(\partial^\alpha\delta\phi\right)
	  + \frac{dV}{d\phi}\delta\phi\right),
\]
and its divergence,
\bea
\nabla_\mu \delta T_{(\phi)}{}^\mu{}_\nu &=&
	-a^{-2}\left(\delta\phi^{\prime\prime} 
	+ 2\frac{a^\prime}{a}\delta\phi^\prime 
	+ k^2\delta\phi + a^2\frac{d^2V}{d\phi^2}\delta\phi 
	+ \frac{1}{2}h^\prime\bar{\phi}^\prime\right)\partial_\nu\bar{\phi}
		\nonumber \\
	&{}&\hspace{0.3in}-a^{-2}\left(\bar{\phi}^{\prime\prime}
	+2\frac{a^\prime}{a}\bar{\phi}^\prime
	+a^2\frac{dV}{d\bar{\phi}}\right) \partial_\nu\delta\phi \nonumber \\
\label{eq:div_dphi}
 &=& \lambda \bar{n}\delta_n\left(\partial_\nu\phi\right)
	 + \lambda \bar{n}\left(\partial_\nu\delta\phi\right),
\eea 
where in the last line we used (\ref{eq:EoM_cphi}) and
(\ref{eq:EoM_dphi}).  

The perturbations to the dark matter stress-energy tensor are defined
to be (following \cite{Ma:1995ey})
\[
\delta T_{DM}{}^0{}_0 = -\delta\rho_{DM}
\]
\[
\delta T_{DM}{}^0{}_i = -\delta T_{DM}{}^i{}_0 
	= \bar{\rho}_{DM} v_i
\]
The other components of the perturbed stress-energy tensor vanish.
The divergence of the stress-energy of the combined dark sector must
vanish to all orders.  Given (\ref{eq:div_dphi}), we demand
\[
\nabla_\mu \delta T_{DM}{}^\mu{}_\nu 
	= -\lambda \bar{n}\delta_n\left(\partial_\nu\bar{\phi}\right)
		-\lambda \bar{n}\left(\partial_\nu\delta\phi\right).
\]
The resulting equations of motion for the dark matter perturbations are
\bea
\label{eq:EoM_dDM1}
\delta^\prime_n &=& -\theta-\frac{1}{2}h^\prime  \\
\label{eq:EoM_dDM2}
\theta^\prime &=& -\left(\frac{a^\prime}{a}
	+\frac{\bar{\phi}^\prime}{\bar{\phi}}\right)\theta
	+ \frac{k^2}{a^2} \frac{\delta\phi}{\bar{\phi}},
\eea
where $\theta = ik^jv_j$.

\begin{figure}
\begin{center}
\epsfig{file=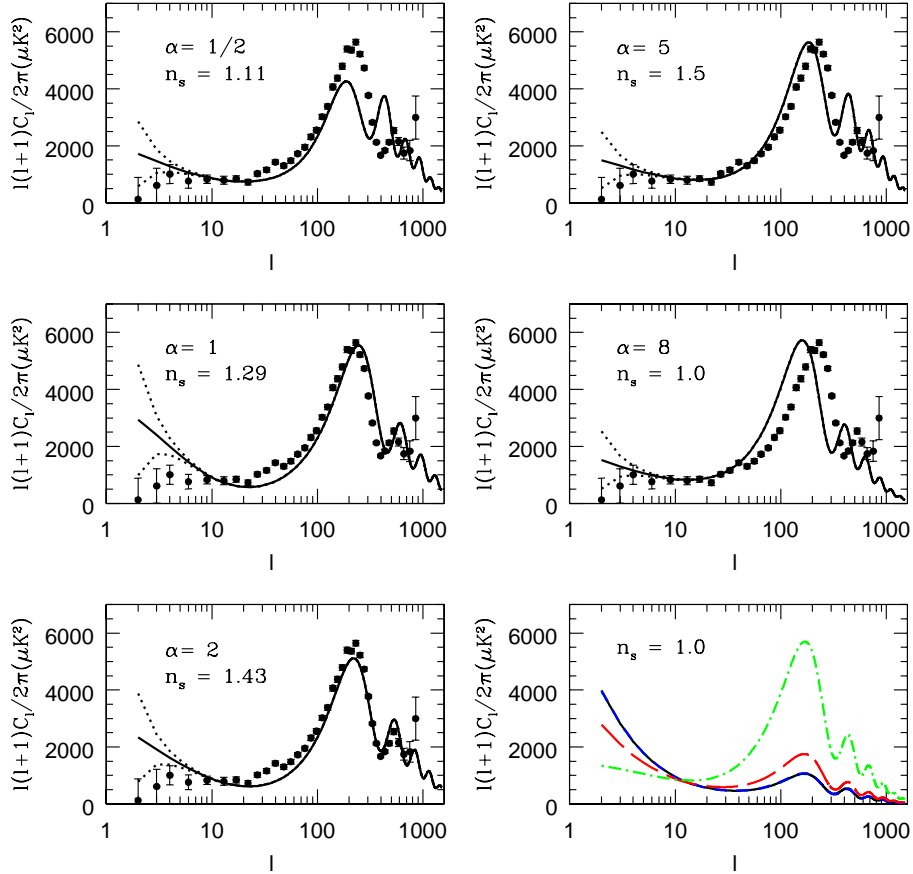, width=5in}
\end{center}
\caption{\label{cmb}
Plots of the CMB temperature anisotropy spectrum for several
models.  In the first five panels, the theoretical spectrum is
compared to the WMAP results \cite{Hinshaw:2003ex}.  For each model,
the best value 
of the scalar spectral index, $n_s$, is shown.  The dotted lines
represent cosmic variance.  The other cosmological
parameters, are set to the best fit values in \cite{Spergel:2003cb}
for illustrative purposes.  In the lower right panel, we plot all five
spectra with the same $n_s=1$, though for $\alpha=1/2,1,2$, the
differences between the spectra are less than the line thickness in
this plot.  The solid line represents these values of $\alpha$, the
long dashed line is $\alpha = 5$, and the dot-dashed line is $\alpha =
8$.}  
\end{figure}

Equations (\ref{eq:EoM_dphi}), (\ref{eq:EoM_dDM1}) and
(\ref{eq:EoM_dDM2}) were integrated along with the standard equations
for baryons and photons using a modified version of CMBFAST
\cite{Seljak:1996is}.  The results are plotted in Fig.~\ref{cmb}.
We were unable to fit any of the models to the WMAP CMB anisotropy
spectrum \cite{Hinshaw:2003ex} with values
of the cosmological parameters that are consistent with other
experiments.  In the first five panels of Fig.~\ref{cmb}, we show
for illustrative purposes
a theoretical CMB temperature anisotropy spectrum with the cosmological
parameters set to the best fit values in \cite{Spergel:2003cb} except for
allowing a variation in the scalar spectral index, $n_s$, a parameter
that describes the initial density power spectrum.  Cosmic variance is
shown as dotted lines, and the error bars on the binned points include
only experimental error.  Effective $\chi^2$ values were calculated for
these models following \cite{Verde:2003ey}, and were all many orders of
magnitude larger than the best fit $\Lambda$CDM model.  In the lower
right panel,  we show all five spectra evaluated with $n_s=1$ in order
to illustrate the dependence of the spectrum on $\alpha$. 

Much of the error in these plots is due to discrepancies between the
theory and the data around the first peak in the spectrum.  One can
imagine, though, that a better fit around the first peak could be
obtained by performing a fit over a larger number of parameters.
Detailed parameter fitting, though, is beyond the scope of this paper.
Even with a better fit around the first peak in the spectrum,
these models face two qualitative problems in predicting the CMB
temperature anisotropy spectrum, one at small scales (large $l$) due
to a low value for the dark matter energy density at last scattering
and the other at large scales (small $l$) due to an enhanced ISW
effect (for a review of CMB physics, see \cite{Hu:2001bc}).  

It is not
surprising that these spectra look like those for a low dark matter
density universe.  In the models considered here, the dark matter
particle mass is increasing with time.  Holding the present dark
matter energy density constant, a larger change in the dark matter
particle mass since last scattering implies a lower value for the
dark matter energy density at last scattering.  At small scales, the
spectra for these interacting models look very much like the spectra
for $\Lambda$CDM models with a low value for $\Omega_{DM}$, models
which also have a low dark matter energy density at last scattering.

The ISW effect, which affects the power in the anisotropy spectrum at
large scales, is due to the integrated evolution of the gravitational
potential along the path of the photons from the last scattering
surface.  During a matter dominated phase, large scale density
perturbations evolve in a manner that keeps the gravitational
potential constant.  Thus, in a universe that is matter dominated from
the time of last scattering to the present, no ISW effect is observed.
During periods when the dominant equation of state parameter, the
ratio of the pressure to the energy density of the dominant energy
component of the universe $w_d$, is changing, the large scale gravitational
potential also changes \cite{Bardeen:kt}.  Thus, the ISW effect is
observed in the $\Lambda$CDM model and in dynamical dark 
energy models due to the recent onset of vacuum domination and the
resulting change in $w_d$ from zero to $-1$ \cite{Coble:1996te}.
In the interacting dark sector models discussed here, the onset of
vacuum domination takes place over a long time, and hence $w_d$ is
also changing over an extended time producing a large ISW effect.
Moreover, the mass of the dark matter particles is increasing with
time, which also causes the gravitational potential to change, and
hence may cause an increase in the observed ISW effect.

The dependence of the anisotropy spectrum on $\alpha$ shown in the
lower right panel of Fig.~{\ref{cmb} is predominantly due to the
dependence of the ISW effect on $\alpha$.  When the scalar field is in
the minimum of its effective potential, the energy density in both
dark matter and dark energy scale as $(a/a_0)^{-3\alpha/(1+\alpha)}$.
This is equivalent to a model with a single fluid with an energy
density evolving as $(a/a_0)^{-3(1+w)}$ with $w=-1/(1+\alpha)$.
For larger values of $\alpha$, $w$ is closer to zero, the value of
$w_{\rm eff}$ during baryon domination.  Hence, larger values of
$\alpha$ imply a smaller change in $w_{\rm eff}$ as the universe
transitions from a baryon dominated epoch to a dark matter/dark
energy dominated epoch, which in turn implies a smaller ISW effect.
In the spectra shown in Fig.~\ref{cmb}, the ISW effect is larger for
the smaller values of $\alpha$.

\section{Conclusions}

The main result of this work is a demonstration of two challenges
facing cosmological models with an interaction between the dark matter
and dark energy.  At the outset we were concerned that an interaction
that causes the mass of the dark matter particles to change in time
would lead to difficulties in reproducing the measured temperature
anisotropy spectrum in the CMB.  In the models considered here, the
mass is a monotonically increasing function of time.  As a result, the
energy density of the dark matter diminishes more slowly in this model
than in the $\Lambda$CDM model, which in turn implies a lower ratio of
dark matter energy density to radiation energy density at the time of
last scattering than in $\Lambda$CDM.  It was conceivable, though,
that since the equations of motion for the perturbations in an
interacting model differ from those in $\Lambda$CDM, some new effect
could compensate for the low dark matter to radiation ratio.  Having
performed the calculation, we now know that this is not the case, and
the initial assumption that the low dark matter to radiation ratio
would be seen in the CMB temperature anisotropy spectrum was correct.

It is not difficult, however, to imagine models that can overcome this
challenge.  One possibility is a model in which the dark matter
particle mass at the time of last scattering is comparable to the
dark matter particle mass today, but varied during the intervening
time.  For example the dark matter particle mass could have been 
decreasing from a large value in the early universe, reached a minimum
at some time between last scattering and today, and be increasing
today.  Such a model could not be ruled out by the considerations in
this work, and detailed study of structure formation in such a model
would be necessary to test it.

Another possibility is to finely tune the initial conditions so that
the scalar field does not evolve enough from the time of last
scattering to today to be ruled out by the CMB data.  In the models
considered here, generic initial conditions lead to the scalar field
sitting in the minimum of its effective potential for a significant
time.  One could finely tune the initial conditions so that the scalar
field does not reach its minimum, and the effective friction due to
the expansion of the universe prevents the field from evolving too
much so that it can satisfy the CMB constraints.  Such a situation,
though aesthetically displeasing, still remains a possibility.  A
similar effect can also be obtained by either introducing a new dark
sector field or a different potential that holds the field near a
fixed value. Ruling out these situations would again require a more
detailed study of structure formation, and this is the approach of 
\cite{Farrar:2003}.

The second challenge to interacting models is avoiding an enhanced ISW
effect at small $l$ in the CMB temperature anisotropy spectrum.  One
of the potential aesthetic benefits of an interacting dark sector
model is that the transition from a dark matter dominated epoch to a
vacuum dominated epoch takes place over an extended period of time
when compared to $\Lambda$CDM, thus softening the coincidence
problem.  Unfortunately, this work has demonstrated that, at least in
the specific model considered here, allowing the dominant effective
equation of state parameter to vary from zero during matter domination
to nearly $-1$ during vacuum domination over too long a time leads to an
enhanced ISW effect that is in conflict with observation.  The methods
discussed above to overcome the challenge of a time varying dark
matter particle mass would likely also overcome this challenge as
well.  Such models allow for only small deviations from $\Lambda$CDM
and more standard dynamical dark energy models, and hence the
evolution of the effective equation of state parameter would be
similar to $\Lambda$CDM and not differ from what is observed.

In this work we have used the CMB temperature anisotropy spectrum to
rule out a class of cosmological models with an interaction between
the dark matter and dark energy.  These models, however, are only
among the simplest possibilities for such an interaction.  More
complicated models may well overcome the challenges pointed out here,
and with strict new observational tests of the $\Lambda$CDM model on
the way, it will be useful to continue to study interacting dark
sector models. 

\section*{Acknowledgments}

The author would like to thank Christian Armendariz-Picon, Jennifer
Chen, Wayne Hu, Eugene Lim, Takemi Okamoto, Simon Swordy, Michael
Turner, Tom Witten, and especially Sean Carroll for useful
conversations.  This work was supported by U.S. Dept. of Energy
contract DE-FG02-90ER-40560 and National Science Foundation grant
PHY-0114422 (CfCP).


\begin{thebibliography}{99}

\bibitem{Persic:1995ru}
M.~Persic, P.~Salucci and F.~Stel,
Mon.\ Not.\ Roy.\ Astron.\ Soc.\  {\bf 281}, 27 (1996)
[arXiv:astro-ph/9506004].

\bibitem{Bahcall:1996mt}
N.~A.~Bahcall,
arXiv:astro-ph/9612046.

\bibitem{Primack:1997av}
J.~R.~Primack,
arXiv:astro-ph/9707285.

\bibitem{Freedman:1999yu}
W.~L.~Freedman,
Phys.\ Scripta {\bf T85}, 37 (2000)
[arXiv:astro-ph/9905222].

\bibitem{Turner:1999kz}
M.~S.~Turner,
Phys.\ Scripta {\bf T85}, 210 (2000)
[arXiv:astro-ph/9901109].

\bibitem{Turner:1997de}
M.~S.~Turner,
``The case for Lambda-CDM'', in {\em Critical Dialogues in Cosmology},
N. Turok {\em ed.}, World Scientific (1997)
[arXiv:astro-ph/9703161].

\bibitem{Bennett:2003bz}
C.~L.~Bennett {\it et al.},
arXiv:astro-ph/0302207.

\bibitem{Netterfield:2001yq}
C.~B.~Netterfield {\it et al.}  [Boomerang Collaboration],
Astrophys.\ J.\  {\bf 571} (2002) 604
[arXiv:astro-ph/0104460].

\bibitem{Halverson:2001yy}
N.~W.~Halverson {\it et al.},
Astrophys.\ J.\  {\bf 568}, 38 (2002)
[arXiv:astro-ph/0104489].

\bibitem{Riess:1998cb}
A.~G.~Riess {\it et al.}  [Supernova Search Team Collaboration],
Astron.\ J.\  {\bf 116}, 1009 (1998)
[arXiv:astro-ph/9805201].

\bibitem{Perlmutter:1998np}
S.~Perlmutter {\it et al.}  [Supernova Cosmology Project Collaboration],
Astrophys.\ J.\  {\bf 517}, 565 (1999)
[arXiv:astro-ph/9812133].

\bibitem{Tonry:2003zg}
J.~L.~Tonry {\it et al.},
arXiv:astro-ph/0305008.

\bibitem{Riess:2001gk}
A.~G.~Riess {\it et al.},
Astrophys.\ J.\  {\bf 560}, 49 (2001)
[arXiv:astro-ph/0104455].

\bibitem{Carroll:2000fy}
S.~M.~Carroll,
Living Rev.\ Rel.\  {\bf 4}, 1 (2001)
[arXiv:astro-ph/0004075].

\bibitem{Peebles:2002gy}
P.~J.~Peebles and B.~Ratra,
Rev.\ Mod.\ Phys.\  {\bf 75}, 599 (2003)
[arXiv:astro-ph/0207347].

\bibitem{Padmanabhan:2002ji}
T.~Padmanabhan,
Phys.\ Rept.\  {\bf 380}, 235 (2003)
[arXiv:hep-th/0212290].

\bibitem{Wetterich:fm}
C.~Wetterich,
Nucl.\ Phys.\ B {\bf 302}, 668
(1988).

\bibitem{Ratra:1987rm}
B. Ratra and
P. J. Peebles,
Phys.\ Rev.\ D {\bf 37}, 3406 (1988).

\bibitem{Frieman:1995pm}
J.~A.~Frieman, C.~T.~Hill, A.~Stebbins and I.~Waga,
Phys.\ Rev.\ Lett.\  {\bf 75}, 2077 (1995)
[arXiv:astro-ph/9505060].

\bibitem{Caldwell:1997ii}
R. R. Caldwell, R. Dave and
P. J. Steinhardt,
Phys.\ Rev.\ Lett.\  {\bf 80}, 1582
(1998)
[arXiv:astro-ph/9708069].

\bibitem{Armendariz-Picon:1999rj}
C.~Armendariz-Picon, T.~Damour and V.~Mukhanov,
Phys.\ Lett.\ B {\bf 458}, 209 (1999)
[arXiv:hep-
th/9904075].

\bibitem{Armendariz-Picon:2000dh}
C.~Armendariz-Picon, V.~Mukhanov
and P.~J.~Steinhardt,
Phys.\ Rev.\ Lett.\
{\bf 85}, 4438 (2000)
[arXiv:astro-ph/0004134].

\bibitem{Armendariz-Picon:2000ah}
C.~Armendariz-Picon, V.~Mukhanov and
P.~J.~Steinhardt,
Phys.\ Rev.\ D {\bf 63}, 103510
(2001)
[arXiv:astro-ph/0006373].

\bibitem{Mersini:2001su}
L.~Mersini, M.~Bastero-Gil and P.~Kanti,
Phys.\ Rev.\ D {\bf 64}, 043508 (2001)
[arXiv:hep-ph/0101210].

\bibitem{Caldwell:1999ew}
R.~R.~Caldwell,
Phys.\ Lett.\ B {\bf 545}, 23 (2002)
[arXiv:astro-ph/9908168].

\bibitem{Carroll:2003st}
S.~M.~Carroll, M.~Hoffman and M.~Trodden,
arXiv:astro-ph/0301273.

\bibitem{Sahni:1999gb}
V.~Sahni and A.~A.~Starobinsky,
Int.\ J.\ Mod.\ Phys.\ D {\bf 9}, 373 (2000)
[arXiv:astro-ph/9904398].

\bibitem{Parker:1999td}
L.~Parker and A.~Raval,
Phys.\ Rev.\ D {\bf 60}, 063512 (1999)
[arXiv:gr-qc/9905031].

\bibitem{Melchiorri:2002ux}
A.~Melchiorri, L.~Mersini, C.~J.~Odman and M.~Trodden,
arXiv:astro-ph/0211522.

\bibitem{Maor:2001ku}
I.~Maor, R.~Brustein, J.~McMahon and P.~J.~Steinhardt,
Phys.\ Rev.\ D {\bf 65}, 123003 (2002)
[arXiv:astro-ph/0112526].

\bibitem{Casas:1992}
J.~A.~Casas, J.~Garcia-Bellido and M.~Quiros,
Class.\ Quant.\ Grav.\  {\bf 9}, 1371 (1992)
[arXiv:hep-ph/9204213].
J.~Garcia-Bellido,
Int.\ J.\ Mod.\ Phys.\ D {\bf 2}, 85 (1993)
[arXiv:hep-ph/9205216].

\bibitem{Wetterich:bg}
C.~Wetterich,
Astron.\ Astrophys.\  {\bf 301}, 321 (1995)
[arXiv:hep-th/9408025].

\bibitem{Anderson:1997un}
G.~W.~Anderson and S.~M.~Carroll,
arXiv:astro-ph/9711288.

\bibitem{Bean:2001ys}
R.~Bean,
Phys.\ Rev.\ D {\bf 64}, 123516 (2001)
[arXiv:astro-ph/0104464].

\bibitem{Farrar:2003}
G.~R.~Farrar and P.~J.~Peebles,
arXiv:astro-ph/0307316.

\bibitem{Amendola:1999er}
L.~Amendola,
Phys.\ Rev.\ D {\bf 62}, 043511 (2000)
[arXiv:astro-ph/9908023].

\bibitem{Amendola:2000uh}
L.~Amendola and D.~Tocchini-Valentini,
Phys.\ Rev.\ D {\bf 64}, 043509 (2001)
[arXiv:astro-ph/0011243].

\bibitem{Amendola:2001rc}
L.~Amendola and D.~Tocchini-Valentini,
Phys.\ Rev.\ D {\bf 66}, 043528 (2002)
[arXiv:astro-ph/0111535].

\bibitem{Amendola:2002bs}
L.~Amendola, C.~Quercellini, D.~Tocchini-Valentini and A.~Pasqui,
Astrophys.\ J.\  {\bf 583}, L53 (2003)
[arXiv:astro-ph/0205097].

\bibitem{Comelli:2003cv}
D.~Comelli, M.~Pietroni and A.~Riotto,
arXiv:hep-ph/0302080.

\bibitem{Amendola:2003eq}
L.~Amendola and C.~Quercellini,
arXiv:astro-ph/0303228.

\bibitem{Gasperini:2001pc}
M.~Gasperini, F.~Piazza and G.~Veneziano,
Phys.\ Rev.\ D {\bf 65}, 023508 (2002)
[arXiv:gr-qc/0108016].

\bibitem{delaMacorra:2002tk}
A.~de la Macorra,
arXiv:astro-ph/0212275.

\bibitem{Mangano:2002gg}
G.~Mangano, G.~Miele and V.~Pettorino,
Mod.\ Phys.\ Lett.\ A {\bf 18}, 831 (2003)
[arXiv:astro-ph/0212518].

\bibitem{Doran:2002bc}
M.~Doran and J.~Jaeckel,
Phys.\ Rev.\ D {\bf 66}, 043519 (2002)
[arXiv:astro-ph/0203018].

\bibitem{Khuri:2003hf}
R.~R.~Khuri,
arXiv:astro-ph/0303422.

\bibitem{Zimdahl:2001ar}
W.~Zimdahl and D.~Pavon,
Phys.\ Lett.\ B {\bf 521}, 133 (2001)
[arXiv:astro-ph/0105479].

\bibitem{Zimdahl:2002zb}
W.~Zimdahl and D.~Pavon,
Gen.\ Rel.\ Grav.\  {\bf 35}, 413 (2003)
[arXiv:astro-ph/0210484].

\bibitem{Chimento:2003ie}
L.~P.~Chimento, A.~S.~Jakubi, D.~Pavon and W.~Zimdahl,
Phys.\ Rev.\ D {\bf 67}, 083513 (2003)
[arXiv:astro-ph/0303145].

\bibitem{Amendola:2002kd}
L.~Amendola,
Mon.\ Not.\ Roy.\ Astron.\ Soc.\  {\bf 342}, 221 (2003)
[arXiv:astro-ph/0209494].

\bibitem{Ma:1995ey}
C.~P.~Ma and E.~Bertschinger,
Astrophys.\ J.\  {\bf 455}, 7 (1995)
[arXiv:astro-ph/9506072].

\bibitem{Seljak:1996is}
U.~Seljak and M.~Zaldarriaga,
Astrophys.\ J.\  {\bf 469}, 437 (1996)
[arXiv:astro-ph/9603033].


\bibitem{Hinshaw:2003ex}
G.~Hinshaw {\it et al.},
arXiv:astro-ph/0302217.

\bibitem{Spergel:2003cb}
D.~N.~Spergel {\it et al.},
arXiv:astro-ph/0302209.

\bibitem{Verde:2003ey}
L.~Verde {\it et al.},
arXiv:astro-ph/0302218.

\bibitem{Hu:2001bc}
W.~Hu and S.~Dodelson,
Ann.\ Rev.\ Astron.\ Astrophys.\  {\bf 40}, 171 (2002)
[arXiv:astro-ph/0110414].

\bibitem{Bardeen:kt}
J.~M.~Bardeen,
Phys.\ Rev.\ D {\bf 22}, 1882 (1980).

\bibitem{Coble:1996te}
K.~Coble, S.~Dodelson and J.~A.~Frieman,
Phys.\ Rev.\ D {\bf 55}, 1851 (1997)
[arXiv:astro-ph/9608122].

\end{thebibliography}
\end{document}